\PassOptionsToPackage{pagebackref,bookmarks}{hyperref}
\documentclass[preprint,journal]{vgtc}                     

\usepackage{amsmath,amsfonts}
\usepackage[table,xcdraw,dvipsnames,svgnames]{xcolor}
\usepackage[caption=false,font=normalsize,labelfont=sf,textfont=sf]{subfig}
\usepackage{textcomp}
\usepackage{adjustbox}
\usepackage{pifont}
\usepackage{dcolumn}
\usepackage{siunitx}
\usepackage{stfloats}
\usepackage{url}
\usepackage{verbatim}
\usepackage{graphicx}
\usepackage{orcidlink}
\usepackage{cite}
\usepackage{outlines}
\usepackage{wrapfig}
\usepackage{mathptmx}                  
\usepackage{ragged2e}
\usepackage{tabu}
\usepackage{multirow}
\usepackage{soul}
\usepackage{dirtytalk}
\usepackage{tabularx}
\usepackage{array}

\onlineid{1695}

\vgtccategory{Research}

\vgtcpapertype{Theoretical/Empirical}

\title{Correcting Misperceptions at a Glance: \break Using Data Visualizations to Reduce Political Sectarianism}

\author{%
  \authororcid{Douglas Markant}{0000-0003-0568-2648},
  \authororcid{Subham Sah}{0009-0007-9446-6731},
  \authororcid{Alireza Karduni}{0000-0001-9719-7513},
  \authororcid{Milad Rogha}{0000-0002-1464-2157},
  \authororcid{My Thai}{0000-0003-0503-2012},
  \authororcid{Wenwen Dou}{0000-0003-0319-9484}
}

\authorfooter{

  \item Douglas Markant, Subham Sah, Milad Rogha, and Wenwen Dou are with the University of North Carolina at Charlotte. Email: \{dmarkant, ssah1, mrogha, wdou1\}@charlotte.edu.
  \item
  	Alireza Karduni is with Simon Fraser University.
  	E-mail: akarduni@sfu.ca.
  \item
  	My Thai is with the University of Florida.
  	E-mail: mythai@ufl.edu.
}

\abstract{%
Political sectarianism is fueled in part by misperceptions of political opponents: People commonly overestimate the support for extreme policies among members of the other party. These misperceptions inflame partisan animosity and may be used to justify extremism among one’s own party. Research suggests that correcting partisan misperceptions---by informing people about the actual views of outparty members---may reduce one’s own expressed support for political extremism, including partisan violence and anti-democratic actions. However, there remains a limited understanding of how the design of correction interventions drives these effects. The present study investigated how correction effects depend on different representations of outparty views communicated through data visualizations. Building on prior interventions that present the \textit{average} outparty view, we consider the impact of visualizations that more fully convey the range of views among outparty members.
We conducted an experiment with U.S.-based participants from Prolific (N=239 Democrats, N=244 Republicans). Participants made predictions about support for political violence and undemocratic practices among members of their political outparty. They were then presented with data from an earlier survey on the actual views of outparty members. Some participants viewed only the average response (Mean-Only condition), while other groups were shown visual representations of the range of views from 75\% of the outparty (Mean+Interval condition) or the full distribution of responses (Mean+Points condition). Compared to a control group that was not informed about outparty views, we observed the strongest correction effects (i.e., lower support for political violence and undemocratic practices) among participants in the Mean-only and Mean+Points condition, while correction effects were weaker in the Mean+Interval condition. In addition, participants who observed the full distribution of out-party views (Mean+Points condition) were most accurate at later recalling the degree of support among the outparty. Our findings suggest that data visualizations can be an important tool for correcting pervasive distortions in beliefs about other groups. 
However, the way in which variability in outparty views is visualized can significantly shape how people interpret and respond to corrective information.  %
\textit{Supplemental materials for this paper are available at this \href{https://osf.io/8crsp/}{OSF repository}}.
}

\keywords{uncertainty visualization, political communication, partisan misperception}

\teaser{
  \centering
  \includegraphics[width=\linewidth]{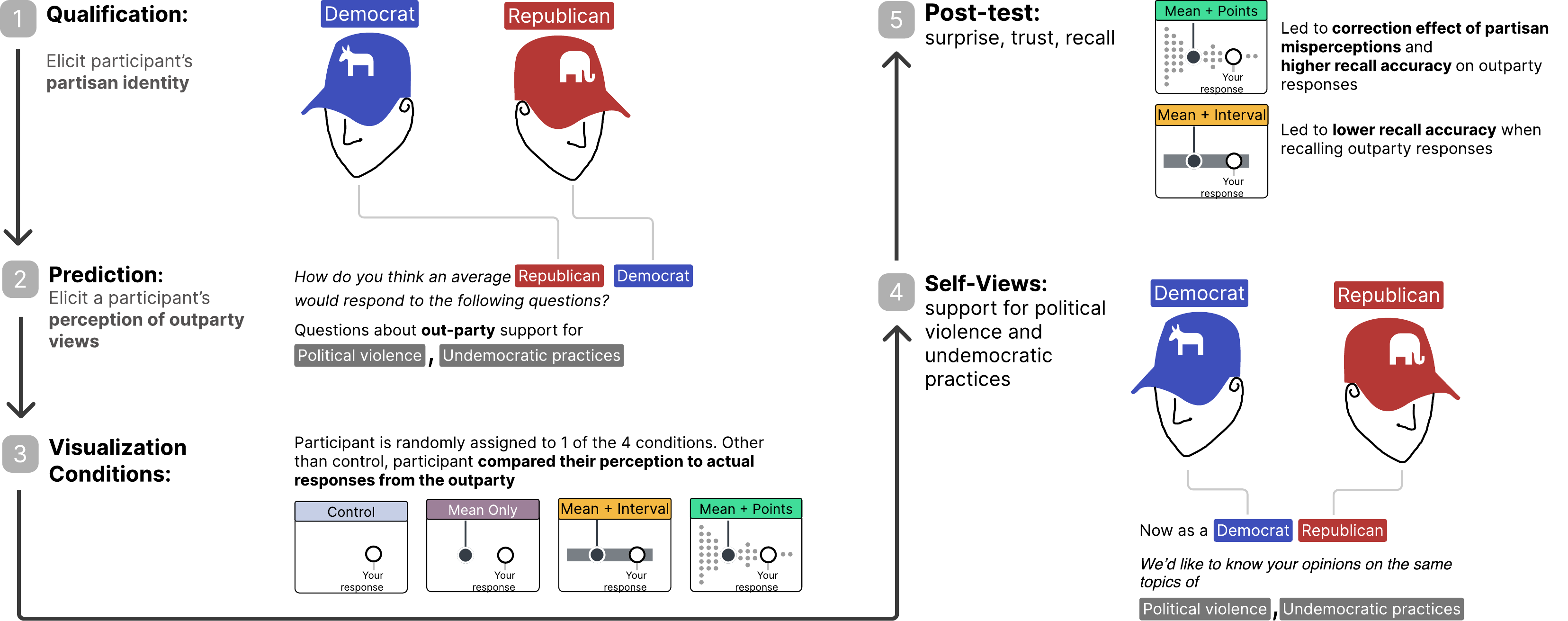}
  \caption{%
  	Our experiment evaluated correction effects of different visual representations of outparty views on one's own support for political extremism. The study included 5 steps, with the overall design and findings annotated in the figure. %
  }
  \label{fig:teaser}
}

\graphicspath{{figs/}{figures/}{pictures/}{images/}{./}} 

\begin{document}



\maketitle


\definecolor{Control}{HTML}{909AB5}
\definecolor{Mean}{HTML}{9F427F}
\definecolor{Interval}{HTML}{E09E19}
\definecolor{Points}{HTML}{10AA67}
\definecolor{Republican}{HTML}{B53835}
\definecolor{Democrat}{HTML}{3E4FBB}

\newcommand{\rev}[1]{\textcolor{black}{#1}}

\newcommand{\Control}{\textbf{\textcolor{Control}{Control}}\space}
\newcommand{\MeanOnly}{\textbf{\textcolor{Mean}{Mean-Only}}\space}
\newcommand{\MeanPoints}{\textbf{\textcolor{Points}{Mean+Points}}\space}
\newcommand{\MeanInterval}{\textbf{\textcolor{Interval}{Mean+Interval}}\space}

\newcommand{\Controlq}{\textbf{\textcolor{Control}{Control}}}
\newcommand{\MeanOnlyq}{\textbf{\textcolor{Mean}{Mean-Only}}}
\newcommand{\MeanPointsq}{\textbf{\textcolor{Points}{Mean+Points}}}
\newcommand{\MeanIntervalq}{\textbf{\textcolor{Interval}{Mean+Interval}}}

\section{Introduction}

Political sectarianism is increasingly recognized not just as a threat to democratic institutions, but as a force that fractures societies and harms individual health and well-being~\cite{van_bavel_political_2024,Fraser_pnasnexus,Finkel_science_2020}.
Evidence suggests that, while political polarization has increased by some measures, people's \textit{perception} of the extent of partisan polarization is far more extreme than warranted \cite{lees2021understanding, wilson_polarization_2020}.
\textit{False polarization} refers to the phenomenon that people perceive the views of other groups to be more extreme (and less diverse) than they actually are \cite{westfall_perceiving_2015}. For instance, \textcolor{Democrat}{Democrats} overestimate the proportion of \textcolor{Republican}{Republicans} who express support for political violence (and vice versa).
False polarization can sustain, and potentially accelerate, actual ideological and affective polarization, as it can cause extreme actions to be seen as more representative of the out-party, be used to justify more extreme actions by one's in-party, and lessen the perceived prospects of compromise~\cite{braley_why_2023,pasek_misperceptions_2022}.

\textbf{Motivation.} Past work has sought to correct false polarization by exposing gaps between perceived and actual views~\cite{dias_correcting_2024,druckman_correcting_2023,landry_reducing_2023,mernyk_correcting_2022,voelkel2024megastudy}.
Correction interventions typically first ask individuals to make a prediction about an out-party view.
Respondents are then shown the actual level of support expressed by members of the other party according to previously collected data.
These corrections have been shown to reduce the perception gap about the other party~\cite{braley_why_2023}.
\rev{Interestingly,} the corrections also reduce respondents' \textit{own} expressed extremism when then asked whether they support the same positions~\cite{braley_why_2023, druckman_correcting_2023}. 
This has led to the suggestion that correcting partisan misperceptions can disrupt the cycle of hyperpartisanship and reinforce mainstream political norms.

\rev{Recent work has highlighted some important limitations of these interventions.}
While they typically correct misperceptions about other groups' views, some studies have failed to show effects on participants' own attitudes~\cite{dias_correcting_2024}.
Correction effects from interventions may also be easily undermined.
For instance, \cite{druckman_correcting_2023} found that following a correction with a caveat about the inherent uncertainty of opinion polling led to the correction having no effect.
Thus there are open questions about the robustness of correction effects and how they are driven by different aspects of the intervention design.

A more fundamental limitation of past work in this area concerns \rev{a nearly exclusive} focus on the \textit{average or most common view} of the out-party, while omitting information about the variability or distribution of views.
Corrections that only communicate the central tendency may obscure \rev{important information about the outparty, such as} the smaller but concerning prevalence of extreme views across the political spectrum.
For example, \cite{mernyk_correcting_2022} found that approximately 5-10\% of respondents in both political parties expressed some willingness to engage in political violence.
A correction that only shows the central tendency \rev{(e.g., low average support for violence among the outparty)} may leave participants with a distorted understanding of outparty views, repeating the error of other forms of political communication by using generic statements about opposing political groups~\cite{peters_how_2021}.

Alternatively, \rev{not communicating the range of outparty views may make it more difficult to reconcile a correction with well-established beliefs about the outparty.}
If people are frequently exposed to examples of extremism on the other side \rev{(e.g., due to hyperpartisanship in media coverage and among political elites \cite{wilson_polarization_2020})}, a focus on an average outparty view that sharply contradicts that prior experience could lead to distrust and rejection of the correction.
\rev{This idea is consistent with misinformation research showing that simply refuting a false belief is less effective than providing alternative explanations or contextual details that help people make sense of conflicting evidence~\cite{lewandowsky_misinformation_2012}.}
\rev{Communicating variability may allow people to realize that while extreme views are indeed present in the outparty, they are far less common than their experience would suggest.}

\textbf{Present study.} These limitations point to a need to consider the effects of including information about the variability or distribution of views (in addition to the average view) when correcting partisan misperceptions.
The present study examines the use of data visualizations in correcting misperceptions about the other party.
Data visualizations offer an efficient means for communicating more complex information about variability in viewpoints.
In addition, visual formats may be especially effective for a correction intervention because they make more salient the gap between a participant's prediction and the data~\cite{rogha_impact_2024, Kim_Gap_2017} as compared to textual comparisons involving numbers.
They may also be more engaging or memorable compared to textual data~\cite{Borkin_visMemorable_2013, yang_trust_2024}.

Our main contributions are the following.
We evaluate the use of data visualizations for reducing politically divisive views, focusing specifically on how correcting misperceptions about the opposing party affects participants' own attitudes toward political violence and undemocratic practices.
Our main research question centers on the impact of visual uncertainty representations which communicate information about the variability of outparty views.
We examine whether uncertainty representations lead to different correction effects, \rev{which are defined (as in prior studies) as the difference in expressed support for extreme actions relative to a control condition that is not exposed to any data about the other party.}
We also expand on earlier work by testing participants' comprehension of the distribution of outparty views by later asking them to estimate both the central tendency and the proportion of extreme views following the intervention.
We also explore the role of perceived surprise and perceived credibility/trust of the data as factors that mediate correction effects.

\textbf{Key Findings.} Consistent with prior research \cite{druckman_correcting_2023}, we found participants dramatically overestimated the support for political violence and undemocratic practices among outparty members. 
Among the visual interventions evaluated, we found: 

\begin{itemize}
\item The \MeanPoints condition (showing both the average and the full distribution of outparty views) led to the \textit{largest correction effect} on participants self-reported support for political violence and undemocratic practices \rev{relative to a \Control condition without any correction.}
The \MeanPoints condition was also associated with the highest accuracy in recalling aspects of outparty response data in a post-test. 

\item The \MeanOnly condition (showing just the average) \rev{led to lower self-reported support for political violence and undemocratic practices compared to the \Control condition, but these effects were weaker than the \MeanPoints condition. Post-test recall of correction data was at similar levels to the \MeanPoints condition.}

\item In contrast, the \MeanInterval condition (showing the average with a summarized range of views \rev{among a majority of outparty members}) was less effective, \rev{leading to higher support for politically divisive actions and lower accuracy in recalling outparty responses compared to the other correction conditions.}

\end{itemize}

\section{Related Work}

\subsection{Correcting partisan misperceptions of support for violence and undemocratic practices}

Several studies have examined whether correction interventions can impact attitudes toward partisan violence, undemocratic practices, and other political views~\cite{voelkel2024megastudy,mernyk_correcting_2022,druckman_correcting_2023}.
These corrections all involve presenting comparisons between a person's predictions about outparty views and the actual views as measured through separate surveys.
Eliciting a prediction first from the participant requires them to draw on their existing attitudes or knowledge about the outparty.
Comparing the prediction and actual view then provides a direct source of feedback about miscalibrated beliefs \cite{markant_when_2023, rogha_impact_2024}.

Correction interventions have varied in how outparty views are presented.
One factor is the level of precision.
Braley et al.~\cite{braley_why_2023} used corrections that identified the most common outparty response (e.g., ``\textbf{Most} Republicans said they would \textbf{Never} support...''), finding that the intervention led to stronger support for democratic norms.
In a study comparing a wide range of approaches for countering antidemocratic attitudes~\cite{voelkel2024megastudy}, this type of correction was among the few interventions that reduced support for partisan violence and undemocratic practices.
Other corrections that provide more precise quantitative feedback \rev{(e.g., the gap between predicted and actual support on a continuous scale, as in the current study)} have also produced lower support for partisan violence and undemocratic practices~\cite{druckman_correcting_2023,mernyk_correcting_2022}.
This kind of correction provides participants more information about the magnitude of the gap between perceived and actual views.
Mernyk et al.~\cite{mernyk_correcting_2022} found that \rev{correction effects were only seen among participants who made larger errors in predicting outparty views, suggesting that recognizing errors in judgment about the outparty is key to reducing one's own expressed support for extreme positions}.

Corrections have also varied in whether they use persuasive language or techniques.
For instance, in some cases corrections are introduced with authoritative statements that people's beliefs about the other party are often wrong.
In a study of outparty dehumanization, Landry et al.~\cite{landry_reducing_2023} used an  expository intervention that informed participants about false polarization without presenting data on specific responses, finding that the intervention reduced outparty dehumanization, partisan animosity, and anti-democratic hostility.

Despite these promising findings, recent studies have raised questions about the impact and robustness of correction interventions~\cite{dias_correcting_2024}.
Some studies have found no effects of corrections on support for undemocratic practices~\cite{dias_correcting_2024} or that corrections reduce partisan animosity without changing support for violence or undemocratic practices~\cite{voelkel_interventions_2022}.
A study by Druckman~\cite{druckman_correcting_2023} suggested that correction effects can also be easily undermined: Correction effects disappear when corrections are followed by information highlighting the inherent uncertainty of survey data or the existence of other, conflicting evidence.
\rev{These mixed findings motivate further investigation into how the design and presentation of correction interventions shape their persuasiveness.}

\subsection{Uncertainty representations and persuasive data visualization}

\paragraph{Uncertainty visualization.} Existing literature has shown that different visual uncertainty representations have important consequences for how visualizations are interpreted~\cite{Padilla_survey_2021}. 
Popular representations include summary plots (e.g. interval plots, mean plots) \cite{padilla_mfv_2023, castro_1Duncertainty_2022}, distributional plots (e.g. quantitle dot plots \cite{Kay_whenismybus_2016}, density plots \cite{Kale_effectSize_2021}), and animated uncertainty representation such as HOPS \cite{Hullman_HOP_2015} and Plinko \cite{yang_swaying_2023}. 
Summary plots display uncertainty in an abstract manner (e.g. an interval that captures most data values) while distributional plots encode the frequency of different values.
Castro et al. found distributional annotations consistently outperformed summary plots in a resource allocation task \cite{castro_1Duncertainty_2022}, and Correll and Gleicher found that violin and gradient plots (i.e. distributional plots) are better than bar charts with error bars (i.e. summary plots) for inferential tasks \cite{Correll_errorBars_2014}. 
However, the ranking of these uncertainty representations may not hold in other tasks or contexts. For example, judgments about probabilities are similar across error bars and violin plots \cite{Hullman_HOP_2015}. Interestingly, Fernandes et al. found that a hybrid density plot overlaid with interval plot outperforms either encoding alone in a transportation decision task \cite{Fernandes_uncertainty_2018}. In addition to visualizing distributions, studies using visualizations of data about different social groups (such as number of deaths or hourly wage) have found that emphasizing variability with a visual encoding can reduce stereotyping in public health and social outcome communication \cite{holder_dispersion_2023, holder_must_2024}.

\paragraph{Uncertainty visualization and partisanship.} In the context of partisanship in the United States and the 2022 US midterm elections, Yang et al. \cite{yang_swaying_2023} found that visualizations encoding probabilistic election forecasts can increase trust and and affect voter's intentions to participate in elections. 
In a related study, Yang et al. \cite{yang_trust_2024} showed that text summaries and quantile dotplots fostered the highest trust over time in a hypothetical election scenario. 

One recent study by Tartaglione and de-Wit~\cite{tartaglione_how_2025} compared how different visualizations impact perceptions about a political outparty.
They presented US and UK participants visualizations comparing the positions of opposing political groups on the topic of immigration.
Their conditions compared bar charts with a full y-axis, bar charts with a truncated y-axis (which magnified intergroup differences), and an icon array histogram constructed to show the degree of overlap in positions across the two groups.
Compared to a control condition, non-truncated bar charts and icon array histograms were associated with decreases in perceived polarization and increases in the perceived similarity of the parties and the likelihood of compromise.

Informed by prior work on uncertainty representations in data visualization, we sought to evaluate the real-world impact of uncertainty visualizations on correcting partisan misconceptions. 
Our work builds on a previous study on correcting partisan misconceptions \cite{druckman_correcting_2023} \rev{in which highlighting uncertainty about the data undermined the correction effect.} Our study includes two visualization conditions that \rev{communicate the variability in outparty views} in a summary and distributional manner. The present work \rev{offers a novel investigation of} how visual interventions with and without uncertainty communication may be used to correct misperceptions of the other political party.   

\begin{figure}[ht!]
  \centering 
  \includegraphics[width=.47\textwidth]{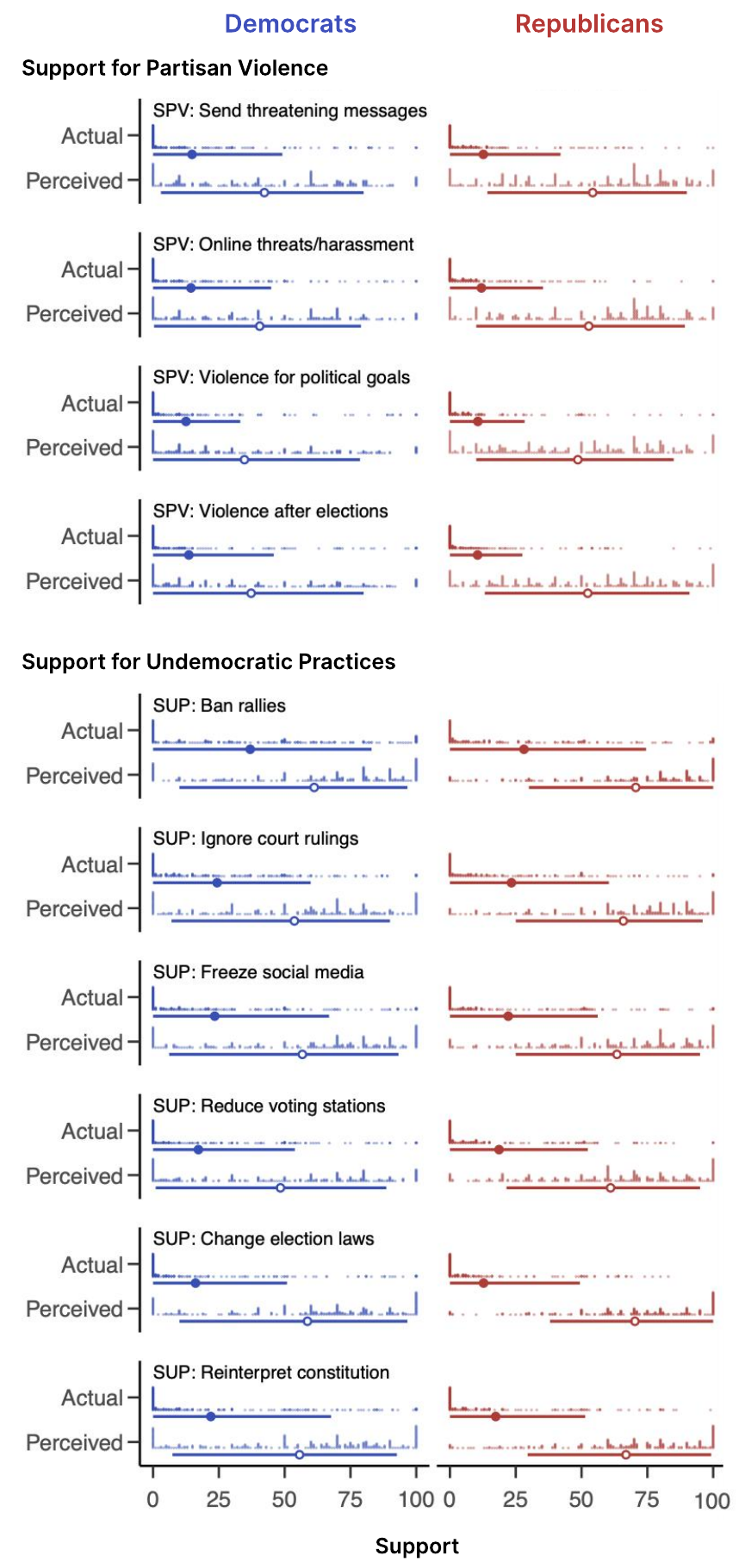}
  \caption{Actual and perceived support for each item. Top row of each panel indicates responses from \textcolor{Democrat}{Democrats} (left) or \textcolor{Republican}{Republicans} (right), while the bottom row of each panel shows predictions of participants from the other party. Dotplots represent the distribution of responses. Intervals indicate the middle 75\% of responses; points indicate the mean.}
  \label{fig:perceptions_by_item}
\end{figure}

\section{Research questions and hypotheses}

Druckman found that the presence of a statement on uncertainty of survey methods or mention of competing surveys led to the disappearance of corrective effects \cite{druckman_correcting_2023}. 
This finding inspired our primary research question: \textbf{How does communicating variability in outparty views (in addition to the average view) impact the correction of partisan misperceptions?} Informed by prior work \cite{Padilla_survey_2021, yang_swaying_2023, yang_trust_2024, Kale_effectSize_2021, holder_dispersion_2023, holder_must_2024}, we hypothesized that including uncertainty representations will \rev{reduce the magnitude of correction effects relative to an intervention that only communicates the average outparty view.} 

Our first two a priori hypotheses concern the effect of experimental condition on self-views related to \textbf{support for partisan violence (SPV)} and \textbf{support for undemocratic practices (SUP)}. Hypotheses were tested independently for each category (SPV and SUP).

\begin{itemize}
\item \textbf{H1a.} SPV/SUP will be lower in the \MeanOnly condition compared to the \Control condition (replication of finding from \cite{druckman_correcting_2023}).
\item \textbf{H1b.} SPV/SUP will be lower in the \MeanOnly condition compared to conditions that include information about the distribution of responses (\MeanInterval and \MeanPoints conditions).
\end{itemize}

\rev{We also investigate \textbf{how people make sense of corrections that conflict with their existing beliefs about the opposing party.}}
Among the conditions that present data about outparty views, we hypothesize differences in the perceived surprise and trustworthiness of the data:

\begin{itemize}
\item \textbf{H2a.} Perceived surprise will be higher in the \MeanOnly condition compared to conditions that communicate the distribution of responses (\MeanInterval and \MeanPoints conditions).
\item \textbf{H2b.} Perceived trustworthiness/credibility of the data will be lower in the \MeanOnly condition compared to conditions that communicate the distribution of responses (\MeanInterval and \MeanPoints conditions).
\end{itemize}

\rev{Lastly, the study included a novel post-test in which participants were asked to recall aspects of the correction data. We use the accuracy of post-test responses to \textbf{explore whether certain visualizations led to better understanding of the outparty views conveyed by the data.}}

The preregistration that includes our study design, hypotheses and analysis methods can be found on \rev{\href{https://osf.io/swt4c/}{OSF}}.

\section{Study design}


To evaluate the role of visualizations in correcting partisan misperceptions and make our results directly comparable with \cite{druckman_correcting_2023}, we leveraged the same questions to assess support for political violence \rev{(SPV; four items)} and support for undemocratic practices \rev{(SUP; six items)}.
As shown in Figure~\ref{fig:teaser}, the study consisted of \textbf{five sequential steps}:

\begin{itemize}
    \item \textbf{Step 1: \rev{Political identity}} – Participants first indicated their political affiliation (Democrat or Republican).
    
    \item \textbf{Step 2: Prediction \rev{of outparty views}} – Participants \textbf{predicted the level of outparty support} for each of the ten political actions.
    
    \item \textbf{Step 3: Visualization conditions} – Participants were shown a \textbf{visual comparison} of their predictions (from Step 2) and the average outparty response, with or without uncertainty representation. Visualization type was a between-subject factor, with each participant randomly assigned to one of four visualization conditions (Figure~\ref{fig:stimuli}). \rev{Participants also answered an open-ended question probing their reactions to the visualizations.}
    
    \item \textbf{Step 4: Self-views} – Participants reported their \textbf{own level of support} for the same 10 actions related to political violence (SPV) and undemocratic practices (SUP). 
    
    \item \textbf{Step 5: Post-test} – Finally, participants report their \textbf{data and visualization perceptions} (e.g., surprise, trust) and completed \textbf{two recall tasks}.
\end{itemize}

The study procedures were approved by the IRB at the University of North Carolina at Charlotte (IRB \#25-0596).
\rev{All study materials are available at the \href{https://osf.io/8crsp/}{OSF repository}.}

\subsection{Participants}

The study was conducted online in March 2025.
A target minimum sample size of 120 participants per condition was determined from an \textit{a priori} power analysis using the data from Druckman~\cite{druckman_correcting_2023}.
The power analysis was performed with G*Power and was based on a oneway ANOVA testing for differences in SPV/SUP among four experimental conditions (alpha = .05, beta = .80, effect size Cohen's f = .16). 
The effect size (f = .16) was based on the observed effect for SUP from~\cite{druckman_correcting_2023}.

$N = 548$ participants were recruited from Prolific.
We used Prolific prescreeners to require that participants  identified as either Democrats or Republicans and had a political ideology of liberal or conservative. 
65 participants were excluded due to one or more preregistered exclusion criteria (33 failed at least one attention check; 28 had a task duration below 5 minutes; 8 indicated a party identification other than Democrat or Republican).
The final sample was comprised of $N=483$ participants ($N=71$ ``weak'' Democrats, $N=168$ ``strong'' Democrats; $N=85$ ``weak'' Republicans, $N=159$ ``strong'' Republicans; 
283 female, 199 male, 1 no sex indicated).
Additional demographic information about the sample is available in Supplementary Table 1.

\subsection{Procedure}

\subsubsection{Political identity and predictions of outparty views} 
Participants first indicated their party identification (\textcolor{Democrat}{Democrat}, \textcolor{Republican}{Republican}, Independent, or other) and political ideology (ranging from \textit{Extremely liberal} to \textit{Extremely conservative}).
\rev{They also indicated their general attitude toward both parties using two ``feeling thermometer'' questions.}
Participants who indicated a party identification other than Democrat or Republican were excluded from analysis.

For each SPV and SUP item, participants were asked to predict the support of a typical outparty member (e.g., Democrats were asked to guess the views of Republicans).
Predictions were made by adjusting a slider on a scale from 0 (no support) to 100 (full support). 
Figure \ref{fig:perceptions_by_item} presents the actual vs. perceived support by political party. 
Similar to the findings from \cite{druckman_correcting_2023}, participants on average \textbf{overestimated their outparty's support} for political violence and undemocratic practices for every question. Detailed results are presented in section 5.1.

\begin{figure}[t]
  \centering 
  \includegraphics[width=.45\textwidth]{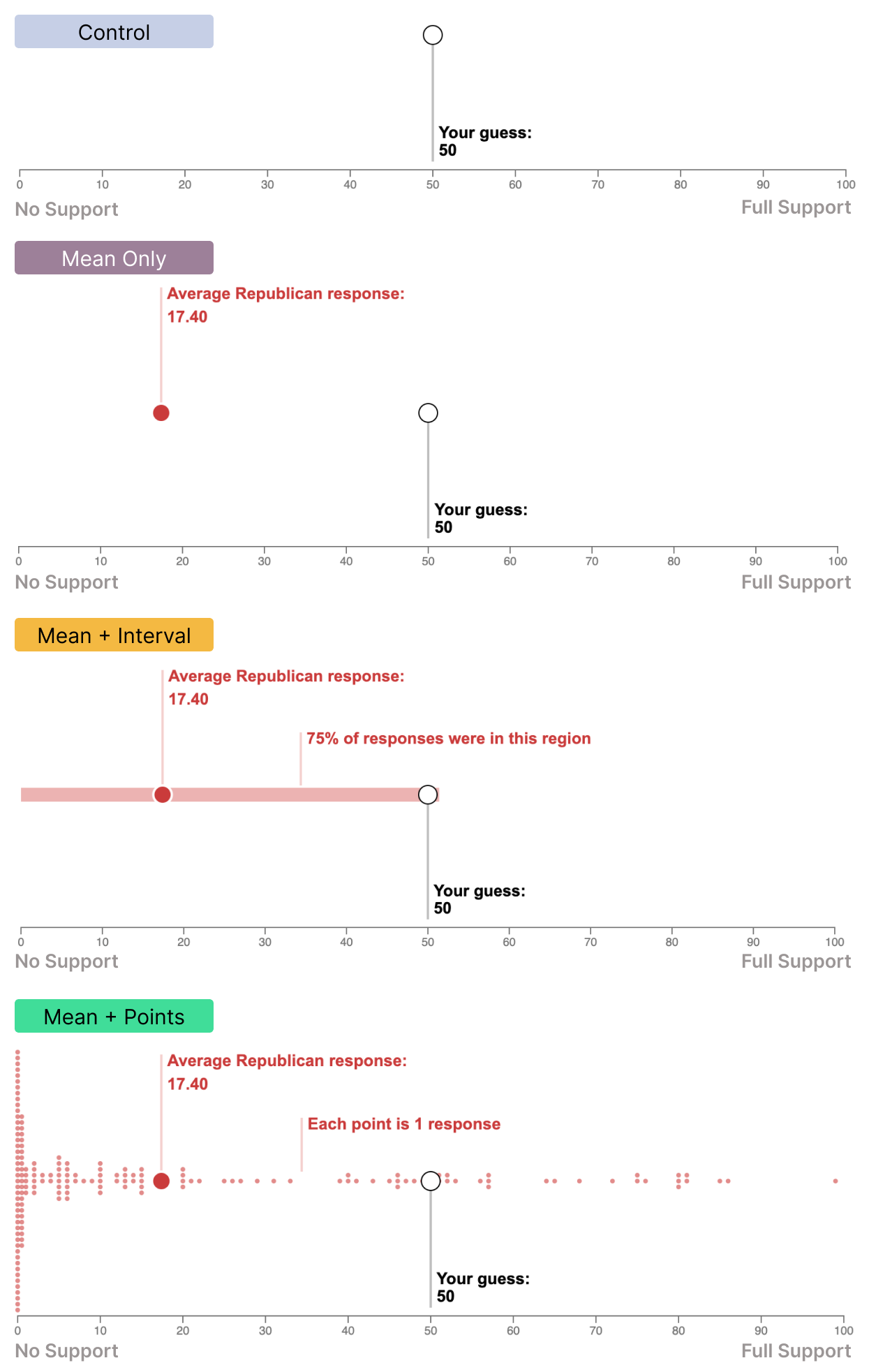}
  \caption{Example stimuli with data about Republican responses that was shown to participants in our study who identified as Democrats.}
  \label{fig:stimuli}
\end{figure}

\subsubsection{Visualization conditions}
There were four between-subjects conditions \rev{(see example stimuli in Figure~\ref{fig:stimuli})}. The \Control and \MeanOnly conditions were similar to the reference study \cite{druckman_correcting_2023}, with the difference being information was shown visually instead of with text. 
All correction conditions used the same baseline survey data from \cite{druckman_correcting_2023}, which included responses from 218 Democrats and 174 Republicans from a survey conducted in June 2022.
\rev{For the correction conditions, participants were instructed that the data was collected in a previous survey of a nationally representative sample of outparty members with the same demographic distribution as the full party. They were not provided any further information about the data source or survey procedures. }

\begin{itemize}
\item \textbf{\Control condition:} Participants were not shown any data about outparty responses to the same 10 questions. Instead, they saw their own predictions plotted on the charts. Participants had to click on each chart to reveal their prediction. 
\item \textbf{\MeanOnly condition:} After revealing their own prediction on a chart, participants clicked a second time to reveal the average response from outparty survey respondents.
\item \textbf{\MeanInterval condition:} Participants saw both the average response and an interval representing the middle 75\% of responses from the outparty respondents. Due to the skewness of the data, all intervals started at 0 but varied in the rightmost position depending on the question (see Figure~\ref{fig:perceptions_by_item}). \rev{The 75\% interval was chosen in order to convey the range of views of a substantial majority of the other party, in keeping with other correction interventions that summarize the position of ``most'' outparty members~\cite{braley_why_2023,voelkel2024megastudy}.}
\item \textbf{\MeanPoints condition:} Participants were shown the average response and the full distribution of responses from outparty respondents. Each dot in this condition represents an individual's response on the response scale.
\end{itemize}

In all conditions, after participants were presented with the visualizations, they were asked to ``summarize the responses that are shown in the charts on this page, including any patterns that stand out to you.''
This open-ended prompt was designed to encourage participants to make comparisons across the 10 questions and, in those conditions where data about outparty views was presented, to make sense of the data without explicitly cuing them to focus on specific visual elements.

\subsubsection{Self-views.} 

After viewing the 10 data visualizations, participants were asked to provide their own self-views for the same 10 questions \rev{using the same response scales from 0 (no support) to 100 (full support)}.
The self-views for SPV and SUP are the main dependent variables in our analysis. 

\subsubsection{Post-test} 

\paragraph{Data perceptions.} Participants in the correction conditions (but not the \textcolor{Control}{Control} condition) then answered four questions about the data that had been presented about the outparty.
One item asked about the degree to which they felt surprised (-3: \textit{Not at all surprised}, +3: \textit{Extremely surprised}).
They were then asked three questions that assessed different aspects of the perceived trustworthiness of the data.
These included judgments about whether the data 1) came from a trustworthy source, 2) was displayed in the visualizations in a way that accurately reflected the survey responses, and 3) was representative of the true views of people in the other party (all on 7 point scales).

\paragraph{Recall tasks.} Finally, participants completed a post-test in which they were asked to recall details about data visualizations that had been presented earlier in the task.
In order to reduce the burden on participants, we selected a subset of five items for the post-test (2 SPV, 3 SUP) which captured the main themes of the topics.
The post-test was not completed by participants in the \textcolor{Control}{Control} condition, as they were not presented with any data from outparty members.

For each item, participants were first asked to recall the average outparty response, which was presented in all three correction conditions. 
They were then asked to indicate the proportion of responses that were above 50, indicating moderate to strong support.
Note that unlike the average response, conditions differed in what information was available to estimate the proportion of support. 
In the \MeanPoints condition, participants could directly observe the proportion of points that fell above 50.
In the \MeanInterval condition, participants could potentially use the upper end of the 75\% interval to \rev{estimate} the proportion of responses above 50.
In the \MeanOnly condition, no information was provided about the distribution of responses.
Note that participants were not instructed to evaluate the proportion of support when the visualizations were shown \rev{in order to adhere to the procedure of \cite{druckman_correcting_2023}}.
We included these questions to explore how people's perceptions of the frequency of extreme views among the outparty were affected by different representations of the data.

\section{Results}

\subsection{Misperceptions of out-party views}

Participants vastly overestimated SPV and SUP among outparty members across all items (Figure~\ref{fig:perceptions_by_item}).
In each question, responses of 0 indicated zero support while 100 indicated full support.
On average, perceived support for partisan violence was overestimated by 40.5 points by \textcolor{Democrat}{Democrats} ($SD = 28.4$) and 24.9 points ($SD = 29.1$) by \textcolor{Republican}{Republicans}. 
Perceived support for undemocratic practices was overestimated by 45.9 points ($SD = 23.8$) by \textcolor{Democrat}{Democrats} and by 32.4 points ($SD = 27.9$) by \textcolor{Republican}{Republicans}. 
Regression models of average errors \rev{(formula: $\texttt{avg\_err} \sim \texttt{condition} + \texttt{party\_id} + \texttt{party\_id\_strength}$)} indicated that overestimation was less severe among \textcolor{Republican}{Republicans} than \textcolor{Democrat}{Democrats} for both SPV ($b = -15.51 \, [-20.70, -10.33]$) and SUP ($b = -13.45 \, [-18.15, -8.75]$). 
There were no differences in the degree of overestimation based on condition or the strength of party identification (see Supplementary Table 2).

\begin{figure}[th!]
  \centering 
  \includegraphics[width=.49\textwidth]{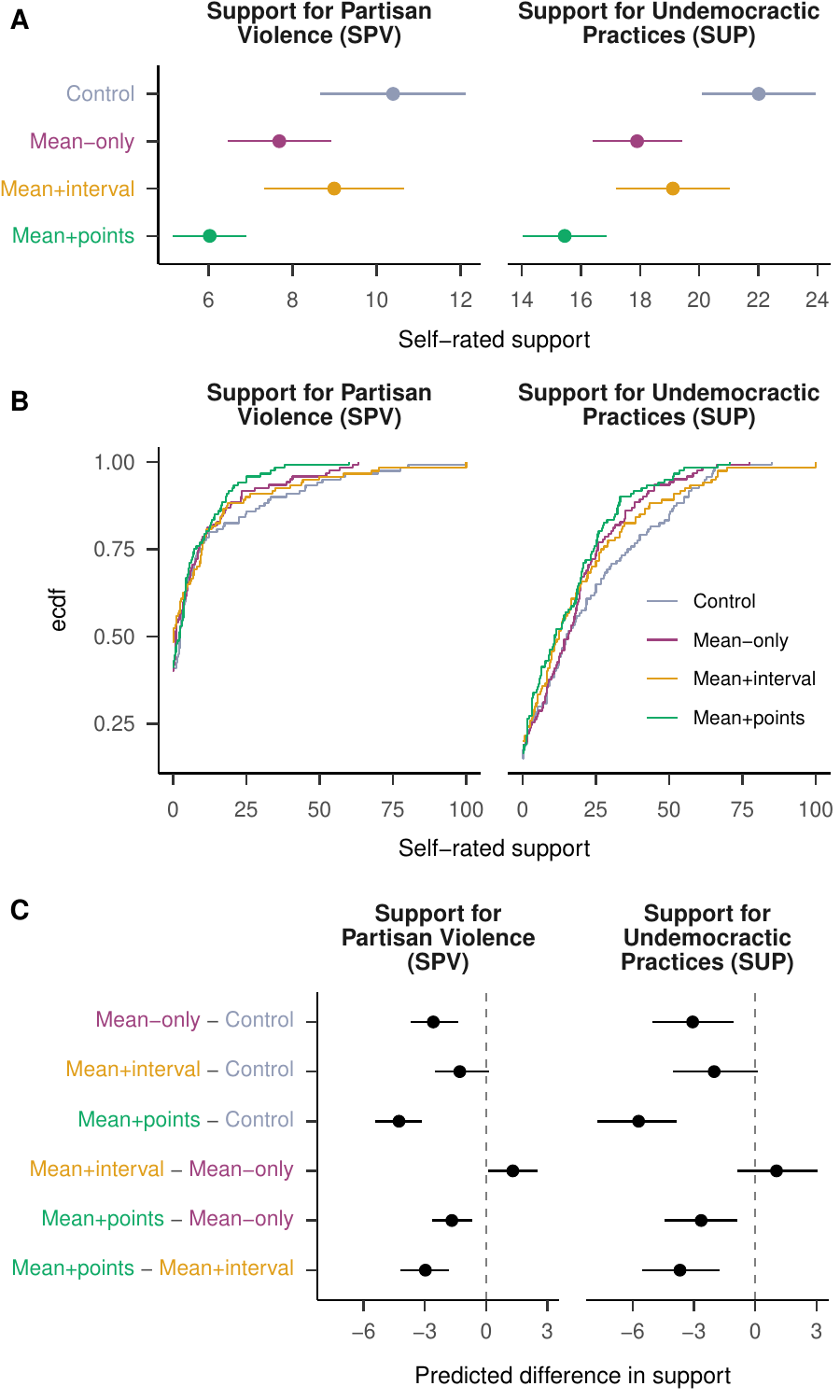}
  \caption{\textbf{A}: Self-rated support (mean and SE) averaged across SPV and SUP items, \rev{on the original response scale ranging from 0 (No support) to 100 (Full support).} \textbf{B:} Empirical cumulative distribution plots for SPV/SUP (averaged across items). The proportion of participants who responded at 0 for all items was similar across conditions (as indicated by similar starting points). Among participants who expressed some support, lower self-ratings were more common in the correction conditions compared to the Control condition (as indicated by steeper rise of curves). \textbf{C:} Pairwise comparisons between experimental conditions \rev{(posterior median and 95\% HDIs)} for predicted SPV and SUP.}
  \label{fig:SPV_SUP}
\end{figure}
\subsection{Correction effects on self-views}

Average self-reported SPV and SUP in each condition are shown in Figure~\ref{fig:SPV_SUP}A.
We first conducted frequentist analyses to compare our findings directly with the previous study that used the same data for corrections~\cite{druckman_correcting_2023}.
A one-way ANOVA for SPV indicated no effect of condition ($F(3,1252) = 1.72$, $p = .16$) and pairwise comparisons with Tukey adjustment indicated no significant differences.
For SUP, there was no overall effect of condition ($F(3,1252) = 1.72$, $p = .16$), while pairwise comparisons indicated that SUP was lower in the \MeanPoints condition than the \Control  condition ($t(479)=2.70$, $p=.04$). 

While the previous study \cite{druckman_correcting_2023} observed a significant correction effect between its mean-only and control conditions, correction effects disappeared in its uncertainty conditions (i.e., when adding uncertainty statements about survey data). 
Interestingly, while our ANOVA analysis found no correction effect in the \MeanOnly condition, the \MeanPoints condition had a significant correction effect relative to the \Control condition. 
In other words, augmenting the mean with a full distribution of outparty responses \rev{was associated with lower self-reported SUP.}

\paragraph{Bayesian multilevel regression analysis.} As per our preregistration, we also used Bayesian multilevel regression to model SPV and SUP, since accounting for individual-level variation and differences between questions may improve our sensitivity to effects.
Models were estimated using the \textit{brms} library~\cite{burkner2017brms} with default priors.
We used zero-one-inflated Beta (ZOIB) regression to model self-rated SPV and SUP.
ZOIB regression models responses (rescaled from 0–1) as a mixture of two distributions: a Bernoulli distribution capturing responses equal to 0 or 1, and a Beta distribution capturing intermediate responses that are greater than zero and less than one. 
ZOIB regression is well-suited to the current data given the highly skewed nature of responses, where a sizable proportion of participants gave responses of 0 for all of the questions.
For results of all Bayesian analyses we report the median and 95\% highest density intervals of the posterior distribution for estimated parameters and contrasts between conditions.

\paragraph{Model specification and parameters.} 
We used the same set of terms to estimate three parameters of the ZOIB model: $\mu$, the mean of the Beta distribution; $\alpha$, the probability of a response being either zero or one; and $\gamma$, among 0/1 responses, the probability of being a 1.
The models included fixed effect terms for individual questions (SPV: 4 items; SUP: 6 items) and experimental condition, as well as random intercepts for participants \rev{(R formula: $\mu, \alpha, \gamma \sim 1 + \texttt{condition} + \texttt{item} + (1|\texttt{pid})$)}.
The \Control condition served as the reference group for terms related to experimental condition.
A fourth parameter which controls the precision of the Beta distribution ($\phi$) was modeled only with an intercept, since additional terms led to a high number of divergences when fitting the model.
The full set of estimated parameters are provided in Supplementary Table 3.

\paragraph{Bayesian modeling results.}
First we focus on the estimated parameters for each condition relative to the \Control group.
For \rev{both SPV and SUP}, for the mean of the Beta distribution ($\mu$) there was a clear negative effect of the \MeanPoints condition relative to the \Control group (\rev{SPV:} $\beta = -.48 \, [-.83, -.13]$; \rev{SUP:} $\beta = -.43 \, [-.65, -.20]$). 
Weaker negative effects were observed for the \MeanOnly condition (\rev{SPV:} $\beta = -.31 \, [-.67, .03]$; \rev{SUP:} $\beta = -.25 \, [-.48, -.03]$), while the direction of the effects for the \MeanInterval was more ambiguous (\rev{SPV:} $\beta = -.11 \, [-.47, .24]$; \rev{SUP:} $\beta = -.12 \, [-.34, .11]$).

For both SPV and SUP, there were no other credible effects of experimental conditions on the parameters governing 0/1 responses ($\alpha, \gamma$), indicating that the proportions of responses at the boundaries of the response scale were similar across conditions.
This is illustrated in the empirical cumulative distribution plots in Figure~\ref{fig:SPV_SUP}B: For both average SPV and SUP, the proportion of zeros is comparable across conditions, as shown by similar starting points for the lines at the left (ranging from 40–48\% of participants for SPV; ranging from 15–20\% of participants for SUP).
However, the steeper rise of the empirical cumulative distribution lines for the correction conditions (especially the \MeanPoints condition) indicates a greater share of lower responses for both self-rated SPV and SUP, consistent with the negative effects on the mean of the Beta distribution governing non-boundary responses. 

To summarize, we observed that, on average, \textbf{all correction conditions reported lower support for partisan violence (SPV) and undemocratic practices (SUP) relative to the \Control condition} (Figure \ref{fig:SPV_SUP}A).
The Bayesian regression modeling indicated \rev{differences between the \Control and correction conditions in the distribution of SPV and SUP responses, with the \textbf{strongest negative effects observed in the \MeanPoints and \MeanOnly conditions}}.
The model results suggest that the correction interventions did not increase the share of participants expressing consistent opposition to all of the SPV or SUP items.
Rather, for those participants who indicated some support, the corrections led to lower expressed support for both kinds of actions.

\paragraph{Comparison of correction effects.}  

\rev{We examined the pairwise differences between all conditions in the predicted correction effects based on the full estimated models (Figure~\ref{fig:SPV_SUP}C).}
Compared to the \Control condition, \rev{consistent} correction effects were observed for the \MeanOnly (SPV: $b = -2.58 \, [-3.73, -1.37]$; SUP: $b = -3.07 \, [-5.04, -1.05]$) and \MeanPoints conditions (SPV: $b = -4.26 \, [-5.42, -3.10]$; SUP: $b = -5.72 \, [-7.72, -3.84]$), \rev{while correction effects were notably weaker in the \MeanInterval condition} (SPV: $b = -1.27 \, [-2.62, -.05]$; SUP: $b = -2.03 \, [-4.11, .02]$).
\textbf{Correction effects were also larger in the \MeanPoints condition compared to the \MeanOnly condition} (SPV: $b = -1.67 \, [-2.71, -.69]$; SUP: $b = -2.64 \, [-4.34, -.76]$) \textbf{and the \MeanInterval condition} (SPV: $b = -2.97 \, [-4.15, -1.85]$; SUP: $b = -3.70 \, [-5.62, -1.87]$). 
The effect for the \MeanOnly condition was larger than the \MeanInterval condition for SPV ($b = -1.30 \, [-2.46, -.07]$) but \rev{comparable} for SUP ($b = -1.03 \, [-3.04, .82]$).

Revisiting the findings from \cite{druckman_correcting_2023} and our hypothesis H1a, our Bayesian analysis results for the \MeanOnly condition aligned with the previous finding of a correction, but with a smaller effect. 
More interestingly, while in H1b we hypothesized that the SUP/SPV would be lower in the \MeanOnly condition compared to the other correction conditions, the analysis revealed the largest correction effects in the \MeanPoints condition for both SUP and SPV. 


\subsection{Data visualization perceptions}

After reporting their own self-views, participants in the correction conditions were asked four questions about their perceptions of the data visualizations that had been shown earlier.
One item asked about their \textbf{perceived surprise}, while the remaining three asked about different aspects of their \textbf{trust in the data} (whether it originated from a trustworthy source; whether it was communicated in a fair manner; and whether it was representative of the actual views of outparty members).
We first evaluated whether these perceptions differed between correction conditions.
In contrast to H2, one-way ANOVAs for each item indicated there were no significant differences between conditions (Supplementary Table 4).

We next examined how perceptions of the data depended on how closely participants' initial guesses came to the actual average views presented in the visualizations.
We also explored whether these perceptions differed between parties.
Both perceived surprise and perceived trust were related to participants' average error in guessing outparty views (Supplementary Figure 1).
We fit Bayesian linear regression models for each item with party identification, average error, and their interaction as predictors \rev{(formula: $\texttt{rating} \sim 1 + \texttt{party\_id} * \texttt{avg\_err}$)}.
Overall, \textbf{larger errors in predicting outparty views were associated with higher ratings of surprise} (\textcolor{Democrat}{Democrats}: $b = .41 \, [.24, .57]$; \textcolor{Republican}{Republicans}: $b = .34 \, [.19, .49]$).
\textbf{Larger errors were also associated with lower ratings for whether the data source was trustworthy} (\textcolor{Democrat}{Democrats}: $b = -.26 \, [-.41, -.11]$; \textcolor{Republican}{Republicans}: $b = -.21 \, [-.34, -.08]$), \textbf{whether it was accurately presented} (\textcolor{Democrat}{Democrats}: $b = -.28 \, [-.43, -.12]$; \textcolor{Republican}{Republicans}: $b = -.19 \, [-.32, -.05]$), and \textbf{whether the data was representative of the true views of the outparty} (\textcolor{Democrat}{Democrats}: $b = -.44 \, [-.59, -.29]$; \textcolor{Republican}{Republicans}: $b = -.21 \, [-.34, -.08]$).
For the final item regarding the representativeness of the data there was a significant interaction between average error and party ID indicating a weaker effect among \textcolor{Republican}{Republicans} ($b = 6.00 \, [.71, 11.14]$).

In sum, reactions to errors in guessing outparty views were generally consistent across conditions and parties.
Larger errors were associated with both greater surprise and lower ratings of trustworthiness, with \textcolor{Democrat}{Democrats} especially likely to question whether the data represented the true views of \textcolor{Republican}{Republicans}.
Interestingly, perceived surprise was \rev{only weakly correlated} with the perceived trustworthiness of the source ($r = -.08 \, [-.18, .02]$), the perceived presentation accuracy ($r = -.06 \, [-.16, .05]$), and the perceived representativeness of the data ($r = -.12 \, [-.22, -.02]$).
This points to substantial variability in how participants responded to the apparent gap between their guesses and the actual outparty views.
While feelings of surprise \rev{may be common when correcting inaccurate beliefs about the other party}, the impact of the correction may be be muted for participants who make sense of that surprise by questioning the credibility of the data.
\rev{In Section \ref{sec:qual} we report an exploratory qualitative analysis intended to gain further insight into the range of participants' responses to the correction interventions.}

\subsection{Post-test accuracy}

\begin{figure}[t]
  \centering 
  \includegraphics[width=.5\textwidth]{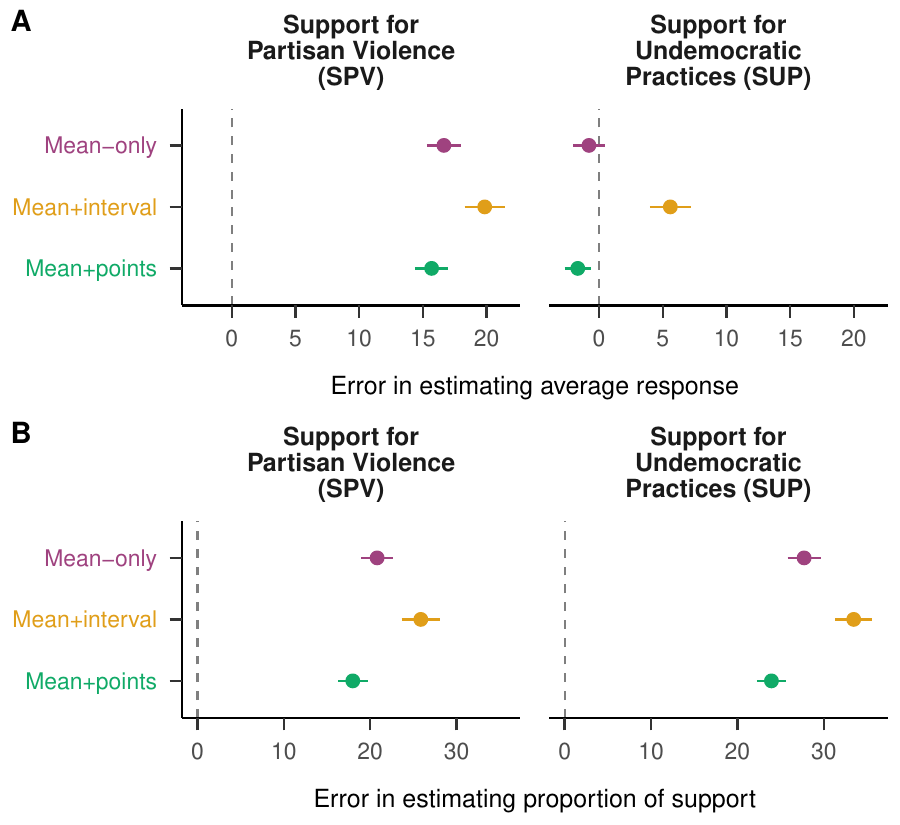}
  \caption{Average error (mean and SE) in the post-test for estimating the average outparty response (\textbf{A}) and the proportion of support (\textbf{B}).}
  \label{fig:posttest}
\end{figure}

Participants who had been shown data were asked to recall aspects of the survey data they saw earlier for a subset of questions (2 SPV items, 3 SUP items).
For each question, they were first asked to identify the \textbf{average level of support} that had been shown by members of the outparty. 
They were then asked to indicate the \textbf{proportion of outparty responses above 50}, indicating moderate-to-strong support for an action.
We measured participants' errors in estimating both the average response and proportion of support.
These were averaged across questions for SVP and SUP.

\paragraph{Post-test estimates of the average response.}
Between the pre-test and post-test, estimates of the average response for the selected questions decreased by 13 points for SPV ($SD=30.38$) and 38 points ($SD = 31.2$) for SUP, indicating that the \rev{\textbf{corrections successfully lowered the perceived outparty support}} on these questions.
In the post-test participants still overestimated the average response for SPV by 17.4 points on average ($SD = 15.61$), while errors for SUP were close to zero ($M = 1.03$, $SD = 14.72$) (Figure~\ref{fig:posttest}A).

We examined whether recall accuracy differed across the three correction conditions.
Differences in average error between conditions was modeled with Bayesian regression with a Student-t family to account for the heavy tailed distribution of errors \rev{(formula: $\texttt{posttest\_error} \sim 1 + \texttt{condition}$)}.
The results for errors in recalling the average response were consistent across SPV and SUP.
Errors were lower in the \MeanPoints condition than the \MeanInterval condition (SPV: $b = -1.51 \, [-2.60, -.47]$; SUP: $b = -3.27 \, [-5.13, -1.43]$).
Errors were also lower in the \MeanOnly condition compared to the \MeanInterval condition (SPV: $b = -1.09 \, [-2.19, -.05]$; SUP: $b = -3.06 \, [-4.95, -1.19]$).
\rev{Recall accuracy for the average response was similar in the \MeanOnly and \MeanPoints conditions} (SPV: $b = -.43 \, [-1.46, .58]$; SUP: $b = -.22 \, [-2.02, 1.58]$).

\paragraph{Post-test estimates of the proportion of support.}
In the post-test participants substantially overestimated the proportion of support by an average of 21.5 points for SPV ($SD = 21.39$) and 28.4 points for SUP ($SD = 21.36$) (Figure~\ref{fig:posttest}B).
We did not include the same judgments in the pre-test to avoid modifying the correction procedure used in prior studies, and therefore can't evaluate whether these perceptions changed as a result of the intervention.

\rev{A Bayesian regression model with the same specification as above was used to evaluate differences in accuracy in estimating the proportion of support.}
\rev{The results indicated a similar pattern across conditions as seen for errors in recalling the average response.}
Errors in estimating the proportion of support were smaller in the \MeanPoints condition than the \MeanInterval condition (SPV: $b = -3.76 \, [-6.77, -.75]$; SUP: $b = -9.12 \, [-14.50, -4.00]$).
\rev{Errors also tended to be smaller in the \MeanOnly condition compared to the \MeanInterval condition, but to a weaker extent} (SPV: $b = -1.63 \, [-4.72, 1.56]$; SUP: $b = -5.23 \, [-10.40, .13]$).
\rev{Recall accuracy for the proportion of support was again comparable in the \MeanOnly and \MeanPoints conditions} (SPV: $b = -2.13 \, [-5.00, .65]$; SUP: $b = -3.96 \, [-9.06, 1.26]$).

In sum, comparing pre- and post-tests, the correction conditions successfully reduced \rev{perceived outparty support for partisan violance (SPV) and undemocratic practices (SUP).}  
\textbf{The \MeanPoints condition led to the \rev{highest} accuracy in recalling the average response and proportion of support}, \rev{but similar levels of accuracy were seen in the \MeanOnly condition}. In contrast, \rev{\textbf{participants in the \MeanInterval condition made substantially larger errors in estimating outparty support in the post-test}.}

\section{Insights on correcting partisan misperceptions from open-ended reflections}
\label{sec:qual}

Participants of both parties demonstrated \rev{highly distorted beliefs about outparty views.}
As seen in prior work, presenting evidence of those misperceptions (by comparing participants' guesses to the real data) led to lower self-expressed support for political violence and undemocratic practices. 
However, post-correction evaluations also suggested that confronting participants with their errors often led to lower ratings of the trustworthiness of the data.
In general, there are many factors which might impact whether individuals are receptive to surprising data that are incongruous with their existing beliefs or attitudes.

We explore these factors through a qualitative analysis of participants' responses to an open-ended question that followed the data visualizations in Step 3 of the study (\textit{``Summarize the responses that are shown in the charts on this page, including any patterns that stand out to you''}). 
We reviewed a total of 464 comments with the guiding question: \textit{"What factors influence reactions to data visualizations that correct inaccurate beliefs about the other party?"}


Thematic analysis followed a systematic, collaborative process to categorize and interpret responses. Two co-authors independently reviewed the data to identify key themes—such as general observations, belief shifts, self-reflection on data deviations, emotional reactions, and trust in visualizations. Inter-rater reliability, measured using Cohen’s Kappa, \rev{across all labels and categories showed an initial agreement of 78.5\%.} Annotators resolved discrepancies through discussion, iteratively refining annotations to achieve 100\% consensus. Thematic analysis revealed six major themes: \textit{Self Reflection on Distance from Average} (201), \textit{Source of Bias} (50), \textit{Lack of Trust in Data / Visualization} (56), \textit{Reactions / Emotions} (93), \textit{Out Party Animosity} (114), and \textit{Receptivity to Attitude Change} (199). Definitions of each theme and their codes are in Supplementary Table 5.

\subsection{Comparisons of actual and perceived views}

The majority of comments included comparisons of one's own response to the reality of the data ($N=201$). 
Consistent with the results of our quantitative analyses, most participants acknowledged misperceptions of the outparty. For example, one participant (\MeanPointsq) wrote: \textit{"I was far off how the Republican think, really glad to see this."} 

Not all participant responses were different from the given data. In fact, several participants ($N=41$) mentioned that their responses were close to the given data, at least for a subset of the questions. For example, a participant wrote: \textit{"I think the responses were exactly what I thought and much more closer if not in line with my initial guesses. I notice some democrats wants to limit republicans from voting."}





Aside from \rev{comparisons of predicted and actual views} 
which clearly reflect the given task of the study, we gathered insights into factors that might make participants \rev{more or less receptive to changing their attitudes about the outparty due to the the correction intervention}.

\subsection{What affects receptivity to attitude change?}

Among the comments that touch on the comparison between perceived and actual views, we found (\textbf{$N=83$}) comments that showed improved attitudes towards the outparty. For example, a participant (\MeanPointsq) who identified with the \textcolor{Republican}{Republican} party said: \textit{"I can see that my responses seem to be pretty far off from the normal answers. I seem to have opposite guesses in relation to the average democratic response. That is pretty eye opening and also makes me think a little bit more."} Similarly, a participant (\MeanPointsq) with \textcolor{Democrat}{Democratic} affiliation mentioned:  \textit{"Republicans, on average, are far less radical than I expected. The vast majority aren't radical at all. Only a small minority of republicans support radical views."}

We also identified (\textbf{$N=116$}) participants \rev{across both parties} whose responses reflected a lack of receptivity to correction after seeing the data. For example, a \textcolor{Democrat}{Democratic} participant (\MeanPointsq)  said:  \textit{"My responses are far higher than the ones from polled Republicans. However I do not find the polled information accurate. It does not represent the actions and experiences I have had in my real life. It does not represent the things that are currently going on in this country on a daily basis or in the last few years."} In a similar vein, a \textcolor{Republican}{Republican} affiliated participant (\MeanIntervalq) said: \textit{"I think the average democrat will support these harsh and horrendous measures very easily contrary to what the chart says is the actual fact because when emotions gets involved logic throwed out of the window. So they will react differently."}

Even with these few anecdotes, it is clear that participants \rev{drew from other sources of existing knowledge and feelings about the other party to make sense of the corrections, consistent with a large body of work on attitude change~\cite{bohner_attitudes_2011}.}
Next, we focus on meaningful patterns from individuals' comments after realizing that their existing views were  misaligned with reality, and which appeared to affect their receptivity to attitude change. To do this, we compared comments categorized as "Does not express outparty receptivity" with the comments that "Express outparty receptivity" across the themes of \textit{emotions, sources of bias, lack of trust, and outparty animosity}. 

\subsubsection{I'm surprised, hopeful, and aware of my biases}

Among comments categorized as "Expresses receptivity towards outparty," there is a clear pattern. 
Recognizing that one's belief about the outparty was widely misaligned with reality leads to an emotion of surprise ($N=15$), sometimes along with a feeling of encouragement ($N=5$), which in turn seems to lead participants to elaborate about their political ideology.
For example, a participant (\MeanIntervalq) wrote: \textit{"They seemed to be way lower than I initial assumed, so maybe I'm not giving enough credit to the general conservative versus the people we see online who like to fan the flames and be extreme."}



Like that example, we found several other comments in which participants reflected on reasons for these biases. 
These reasons included mentions of toxic media ($N=7$), the general political climate ($N=4$), and a general acknowledgment of being biased ($N=3$). For example, A \textcolor{Democrat}{Democratic} participant (\MeanPointsq) wrote: \textit{"I can see how the current political climate and all the outrageous things I read every day about the current administration is influencing my opinion of republicans. I was glad to see the averages were much lower than my guesses, though (although I feel like it's probably safe to assume the pro-MAGA, Trump/Elon loving conservatives that frighten me aren't on platforms taking these surveys)."}

Aside from a few comments that included mentions of skepticism towards the data or the sample size, this category of comments did not contain notions of outparty animosity or deep mistrust in the data. 
Looking at the group of comments that "Does not express outparty receptivity," however, paints a different picture.

\subsubsection{I'm angry, the data is wrong}

Emotions and reactions identified within this group of comments were widely different. Although some reflected a sense of surprise ($N=3$), several exhibited anger ($N=4$), fear or disappointment ($N=4$), or concern ($N=5$). Such heightened emotions were present in both \textcolor{Democrat}{Democratic} and \textcolor{Republican}{Republican} comments. For example, a \textcolor{Democrat}{Democratic} participant (\MeanOnlyq) mentioned:  \textit{"I think they are hypocritical if they believe the polar opposite of what they have elected. Maybe these Republicans should join the Democratic party if they support democracy this much."} We observed a similar pattern among \textcolor{Republican}{Republican} individuals with a sense of deep outparty mistrust. For example, a participant (\Controlq) said:  \textit{"In all instances referenced above, of anything that supports the Democrats warped view of "social progress" I chose 100\% because as someone who has been politically aware since high school, the patterns are just so evident. Now that we have conservative leadership, everything the Democrats have been doing, their agenda, is being exposed, another reason why selecting the options I did above."}

Such sentiments were often accompanied by explicit mentions of a lack of trust ($N=26$). Several participants mentioned that they believe the respondents to the survey were dishonest ($N=16$). For example, a participant (\MeanOnlyq) wrote: \textit{"I believe the Republicans lied on their surveys. The stats that are shown and the action I have personally been privy to do not show the same standard. Therefore I believe Republicans are lying about how violent and hate filled they are because they know they will be shunned and ostracized for acting how they actually want."}

As also expressed in the above quote, there were several examples where participants expressed a perceived misalignment from what the data shows to their own experiences ($N=7$) sometimes accompanied by questioning the veracity of the data source ($N=3$). For example a participant (\MeanPointsq) wrote: \textit{"I think many of these charts are off because lots of Democrats are out right now using violence and harassment on Republicans. They are destroying Tesla dealerships to "protest" Musk. They are a far more violent party and see "protesting" as an excuse to violently attack people."} 

The themes that emerged from these open-ended evaluations of the data provide important additional context for understanding why corrections succeed or fail.
While highlighting partisan misperceptions often created feelings of surprise, participants varied widely in how they made sense of the apparent gap between their perceptions and the data. 
These findings suggest that corrections may be more effective when they aid that sensemaking process, e.g. by including explanations for possible sources of bias that lead to false polarization.

\section{Discussion, Limitations, and Future Work}
\label{sec:discussion}

Political sectarianism has been fueled by profound misjudgments about the views of members of opposing political parties~\cite{Finkel_science_2020,westfall_perceiving_2015}.
Existing work suggests that a promising strategy for \rev{tempering partisan animosity and divisiveness is to inform} people about their errors when predicting the views of outpartisans~\cite{voelkel2024megastudy}. However, there are open questions about when and why these corrections work, including what is the best approach for communicating the views of other groups.
While existing studies have focused on conveying the average view of outparty members, such an approach may paint a misleading picture, for example, by obscuring the degree of consensus or the relative prevalence of moderate vs. extreme positions.
This is a particularly important consideration for questions about political extremism where responses tend to be heavily skewed and the central tendency may not be the best representation of the distribution of views in the other party.

\subsection{Comparing visualizations of outparty views}

We examined how data visualizations can be used to communicate outparty views in order to correct partisan misperceptions.
We found that a correction which included the average and the full distribution of responses (\MeanPoints condition) \rev{led to the lowest self-reported support} for partisan violence (SPV) and support for undemocratic practices (SUP).
This result was counter to our expectations that showing the full distribution might draw attention to the most extreme positions (i.e., individuals in the other party who expressed strong support for violent or antidemocratic actions).
In the post-test, participants in the \MeanPoints condition were also the most accurate at \rev{recalling the degree of outparty support conveyed by the data}, further suggesting that including the full distribution did not lead to disproportionate weighting of the relatively small number of extreme responses.
\rev{Rather, it seems likely that} participants in this condition paid more attention to the large cluster of responses at the lower end of the scale indicating ``no support'' for each type of action\rev{---a strongly anti-sectarian contingent that is obscured when only reporting the average response.}
\rev{Thus the \MeanPoints condition may be especially effective for these topics because of the highly skewed distribution of views.}

Our findings for the \MeanInterval condition were more aligned with our hypothesis that visual uncertainty representations would undermine the correction, \rev{as that condition led to smaller correction effects compared to both the \MeanPoints and \MeanOnly conditions}.
In other tasks that expose mistaken prior beliefs using data visualizations, the presence of uncertainty representations has been linked to smaller changes in beliefs~\cite{markant_when_2023}.
Here the \MeanInterval condition led to the smallest correction effects relative to the \Control condition, effects that were also smaller than \MeanOnly and \MeanPoints conditions.
Moreover, this condition was associated with \rev{the largest overestimation errors when later recalling} the average outparty view and the proportion of support (i.e., the proportion of responses above the midpoint of the scale). 
Lower accuracy in recalling the proportion of support is especially notable because the visualizations included a reference point (the upper end of the 75\% interval) that could be used to estimate where 25\% of the responses fell, and in many cases that reference point was close to the midpoint of the scale (see ``Actual'' intervals in Figure~\ref{fig:perceptions_by_item}).
Rather than encoding the absolute extent of the intervals, \rev{we expect that} participants may have focused more on evaluating whether their guesses fell inside them.
Increased attention to their own predictions or the upper end of the interval could have led to better encoding of those quantities, which in turn may have biased later judgments in the post-test.
Observing that their guess fell within an interval may have also been interpreted as being ``correct,'' leading to less motivation to inspect the difference between their estimate and the average response.
\rev{This would be consistent with other evidence that people misinterpret intervals in a deterministic fashion, failing to understand that they represent a range of possibilities with varying degrees of likelihood~\cite{joslyn_visualizing_2021}}.

\rev{Our conclusions are naturally limited by how we chose to summarize outparty views in the \MeanInterval condition. Other work has shown that different interval-based representations of the same data can strongly impact comprehension and decision making~\cite{hofman_how_2020}. We chose to represent the range of views for an intuitively clear majority (75\%) of outparty members. Larger intervals (e.g., 95\%) might further weaken correction effects because they would include more extreme positions without communicating their relatively low frequency. 
Smaller intervals (e.g., 50\%) might have similar effects as the \MeanPoints condition if they draw attention to a smaller range of anti-sectarian positions, but they might also be perceived as less representative of the breadth of views in the outparty.}

Taken together, our results suggest that data visualizations can be an important tool for correcting partisan misperceptions, but that the method for representing outparty views matters.
Including the full distribution of views strengthens the correction effect and leads to better (albeit still biased) understanding of the prevalence of more extreme positions.
In contrast, an interval summarizing the range of views held by a majority of outparty members may be less effective, \rev{consistent with other evidence that summary plots} can lead to poorer comprehension and inference compared to distributional representations~\cite{castro_1Duncertainty_2022,Correll_errorBars_2014}.
\rev{It remains to be seen whether these findings generalize to other kinds of data or visualization design choices.}
\rev{Moreover, while we have speculated that the \MeanPoints and \MeanInterval conditions may involve different patterns of attention and/or reasoning about visual elements, further work is needed to explore these differences in processing and their impact on correction effects.}

\subsection{Improving correction interventions}

Our findings also offer novel insights into a range of factors which impact the effectiveness of correction effects.
It is important to note that correction effects were relatively small, but of a similar magnitude to those found in earlier studies~\cite{dias_correcting_2024}.
Effects were also smaller than the closest comparison study that relied on the same data and general procedure for corrections~\cite{druckman_correcting_2023}.
This attenuation could be due to a number of factors, including procedural differences in how information was presented to participants.

It is also important to consider the political climate in the U.S. at the time of the study (March 2025), a period during which a Republican administration engaged in high-profile actions perceived by many as threatening democratic norms.
Our qualitative analysis of evaluations of the data suggested these contemporaneous events influenced participants' receptivity to the corrections.
While many participants reported feeling surprised after seeing data that was very different from their guesses, there was a notable divergence between participants who responded with self-reflection or hope vs. those who appeared to discount the data by questioning its real-world relevance. 
Some participants made explicit mention of recent events that contradicted the main message of the data visualizations in the qualitative comments.

Future work should aim to better understand how people make sense of corrections in light of other knowledge that informs their judgments about political opponents, and how the trajectory of that sense-making process impacts shifts in their own attitudes.
Correction interventions \rev{with data visualizations} could be designed to support positive attitude changes, e.g. by including annotations that draw attention to specific visual features such as the cluster of low responses.
The data could also be framed with direct statements about false polarization~\cite{braley_why_2023} or explanations for why misperceptions exist (e.g., highlighting the role of biased media coverage---\rev{an explanation that was spontaneously offered by some of our participants}).
Another fruitful direction could be to include comparisons of both outparty and inparty views, which may highlight the commonality in the distribution of views across opposing parties~\cite{tartaglione_how_2025}.
\rev{Finally, since a number of participants appeared to doubt the credibility of the correction data when it contradicted their predictions, additional information could be provided about the data source or survey procedure to bolster its perceived trustworthiness.}





\section{Conclusion}

This study highlights the potential for using data visualizations to correct misperceptions that foster political sectarianism.
We find that it is important to consider how visual representations of outparty views can lead to different outcomes.
Contrary to concerns that communicating variability might draw attention to extreme views, our findings suggest that representing the full distribution of views leads to greater accuracy about outparty views and less support for political violence and undemocratic practices, likely due to increased attention to the predominance of low support in the other party. Our findings point to the need for more targeted and psychologically informed interventions---such as guiding visual attention, narrative framing, or inparty-outparty comparisons---to increase the persuasive impact of correction interventions. Future work could explore how people interpret visual corrections and how design strategies can foster receptivity to positive attitude change and greater political understanding across divides.

\section{Acknowledgements}

The authors thank Dr. James Druckman for providing the baseline survey data from~\cite{druckman_correcting_2023} for use in the correction interventions. The study is supported by National Science Foundation Award \#2323795.

\section{Supplementary Materials}

Supplementary information is available at the \href{https://osf.io/8crsp/}{OSF repository}.

\bibliographystyle{abbrv-doi-hyperref}
\bibliography{template}

\end{document}